\documentstyle[11pt]{article}

\def\s2m{$OSp(1/2m,R)$}
\def\s32{$OSp(1/32,R)$}
\def\s64{$OSp(1/64,R)$}
\def\2m{$Sp(2m,R)$}
\def\32{$Sp(32,R)$}
\def\64{$Sp(64,R)$}
\def\slr{$OSp(1/32,R)_L \times OSp(1/32,R)_R$}
\def\s32l{$OSp(1/32,R)_L$}
\def\s32r{$OSp(1/32,R)_R$}
\def\s32lr{$OSp(1/32,R)_{L-R}$}
\def\arr{\Longrightarrow}

\newcommand{\eq}{\begin{equation}}
\newcommand{\en}{\end{equation}}
\newcommand{\eqn}{\begin{eqnarray}}
\newcommand{\enn}{\end{eqnarray}}
\newcommand{\nn}{\nonumber }
\newcommand{\lwv} { lowest weight vector}

\begin{document}
\begin{titlepage}
\begin{flushright}
PSU-TH-195 \\
  March 1998
\end{flushright}
\begin{center}
{\bf  Unitary Supermultiplets of $OSp(1/32,R)$ and 
M-theory} \\
\vspace{1cm} 
{\bf M. G\"{u}naydin\footnote{Work supported in part by the
National Science Foundation under Grant Number PHY-9631332. \newline 
e-mail: murat@phys.psu.edu}}   \\
\vspace{.5cm}
Physics Department \\
Penn State University\\
University Park, PA 16802 \\
\vspace{1cm}
{\bf Abstract}
\end{center}
We review the oscillator construction of the unitary representations of 
noncompact groups and supergroups and  study  the unitary supermultiplets  of $OSp(1/32,R)$ 
in relation to M-theory. $OSp(1/32,R)$ has a singleton supermultiplet 
consisting of a scalar  and a spinor field. Parity invariance  leads us to consider $OSp(1/32,R)_L \times OSp(1/32,R)_R$ as the "minimal" generalized AdS supersymmetry algebra of
M-theory  corresponding to the embedding of two spinor representations of $SO(10,2)$ in the fundamental representation of $Sp(32,R)$. The contraction to the Poincare superalgebra with central charges proceeds via a diagonal subsupergroup $OSp(1/32,R)_{L-R
}$ which contains the common subgroup $SO(10,1)$ of the two $SO(10,2)$ factors. The parity invariant singleton supermultiplet of
$OSp(1/32,R)_L \times OSp(1/32,R)_R$ decomposes into an infinite set of "doubleton" 
supermultiplets of the diagonal $OSp(1/32,R)_{L-R}$. There is a unique "CPT self-conjugate" 
doubleton supermultiplet whose tensor product with itself yields the  "massless" generalized $AdS_{11}$ supermultiplets. The massless graviton supermultiplet contains fields corresponding to those of 11-dimensional supergravity plus  additional ones. Assu
ming that an AdS phase of M-theory exists we argue that the doubleton field theory must be the holographic  superconformal field theory
in ten dimensions that is dual to  M-theory in the same sense as the duality between the $N=4$ super Yang-Mills in $d=4$ and the $IIB$ superstring over $AdS_5 \times S^5$.  
 
\end{titlepage}
\section{Introduction}
\setcounter{equation}{0}   

In a recent work Maldacena \cite{mald}  conjectured that the large $\cal N$ limits
of certain conformal field theories in $d$ dimensions are dual to supergravity (
and superstring theory in a certain limit) on the product of $d+1$ dimensional 
AdS spaces with certain spheres.  Maldacena's conjecture was motivated by earlier work
on  $p$-branes \footnote{See the references quoted in \cite{mald}.}.
 The prime example of this duality is between the
$ N =4$ super Yang-Mills in $d=4$ and the IIB superstring over $AdS_5 \times S^5$
in the large $\cal N$ limit. In \cite{mgdm} it was pointed out how the conjecture of
Maldacena   can be understood naturally within the framework of  work
done long time ago on Kaluza-Klein supergravity    theories. Referring to \cite{mgdm}
for details and references to the earlier  work we shall summarize briefly the most
relevant results below. Since the work of Maldacena there has been an explosion of 
interest in AdS supergravity theories, p-branes and their relations to superconformal field
theories (\cite{sfcf}-\cite{jgomis}).

In \cite{gw} the unitary 
supermultiplets of the $d=4$ AdS supergroups 
$OSp(2N/4,R)$  were constructed and the spectrum of the $S^{7}$ compactification
of  eleven dimensional supergravity was fitted into an infinite tower of unitary
supermultiplets of $OSp(8/4,R)$. The ultra-short singleton supermultiplet sits
at the bottom of this infinite tower of Kaluza-Klein modes and 
 decouple from the spectrum as local gauge degrees of freedom
\cite{gw}. 
However , even though the singleton supermultiplet 
decouples from the spectrum as local gauge modes, one can 
generate the entire spectrum of 11-dimensional supergravity over $S^7$
by tensoring the singleton supermultiplets repeatedly and restricting oneself to "CPT self-conjugate "
vacuum supermultiplets. \footnote{ As will be explained later a vacuum supermultiplet
corresponds to a unitary representation whose lowest weight vector is the Fock vacuum.}

The compactification of 11-d supergravity over the four 
sphere $S^{4}$ down to seven dimensions was studied in 
\cite{gnw,ptn}.
The spectrum of 11-dimensional supergravity over $S^4$ fall into 
an infinite tower of unitary supermultiplets
of $OSp(8^{*}/4)$ with the even subgroup $SO(6,2)\times USp(4)$ \cite{gnw}.
Again the doubleton supermultiplet of $OSp(8^{*}/4)$ decouples from the  
spectrum as local gauge degrees of freedom. It consists of five scalars, four fermions and
a self-dual two form field \cite{gnw}. The entire physical spectrum of 11-dimensional
supergravity over $S^4$ can be obtained by simply tensoring the doubleton 
supermultiplets repeatedly and restricting oneself to the vacuum
supermultiplets \cite{gnw}.

The spectrum of the $S^{5}$ compactification of ten dimensional
IIB supergravity was calculated in \cite{gm,krv}. 
The entire spectrum  
falls into an infinite tower of massless and massive unitary supermultiplets of
$N=8$ $AdS_{5}$ superalgebra $SU(2,2/4)$ \cite{gm}.
The "CPT self-conjugate" doubleton supermultiplet of 
$N=8$ $AdS$ superalgebra  decouples from the physical 
spectrum as local gauge degrees of freedom. 
By tensoring the doubleton supermultiplet with itself 
repeatedly and restricting oneself to the $CPT$ self-conjugate vacuum
supermultiplets one generates the entire spectrum of
Kaluza-Klein states of ten dimensional IIB theory on $S^{5}$. 

In \cite{gm,mgnm2} it was pointed out that the $N=8$  $AdS_5$ 
doubleton supermultiplet does not have a Poincare limit 
and its field theory exists only on the boundary of $AdS_5$ which
can be identified with the $d=4$ Minkowski space . Hence the doubleton field
theory of $SU(2,2/4)$ is the conformally invariant 
$N=4$ super Yang-Mills theory in $d=4$. 
Similarly, the singleton supermultiplet of $OSp(8/4,R)$ and the doubleton 
supermultiplet of $OSp(8^*/4)$ do not have a Poincare limit in $d=4$ and
$d=7$, respectively, and their field theories are conformally invariant theories
in one lower dimension \footnote{ see \cite{mgdm} for references }. 
Thus we see that the proposal of Maldacena follows directly from the above
mentioned results if we assume that the spectrum of the 
superconformal field theories fall into ("$CPT$ self-conjugate" ) 
vacuum supermultiplets. Remarkably, this is equivalent to assuming that
the spectrum consists of color singlet supermultiplets! 
Furthermore, taking the large $\cal N$ limit allows one to tensor arbitrarily many
doubleton supermultiplets so as to be able to obtain the entire infinite tower 
of Kaluza-Klein states.

Our goal in this paper is to extend these results to the maximal possible dimension.
In particular, we would like to find out if there exists an AdS phase of M-theory in the
maximal space-time dimension
that is dual to some superconformal quantum field theory. We are not able to give a definitive
answer to this question. However, we find  evidence from the supermultiplet
structure of generalized AdS supergroups in 11 dimensions that M-theory can indeed 
have an AdS phase  that is dual to a superconformal
 doubleton field theory in ten dimensions.  In sections 2 and  3 we review briefly the oscillator
construction of the unitary lowest weight representations of noncompact groups and
supergroups. In section 4 we study the unitary lowest weight representations of
the supergroup $OSp(1/2m,R)$  , in general, and $OSp(1/32,R)$ in particular. 
In section 5 we discuss the unitary supermultiplets of $OSp(1/32,R)$ in relation
to M-theory and physics in ten and eleven dimensions. The
parity invariance of M-theory requires the extension of $OSp(1/32,R)$ to a larger
supergroup that admits parity invariant representations. The "minimal" parity
symmetric supergroup is $OSp(1/32,R)_L\times OSp(1/32,R)_R$ .\footnote{That 
this supergroup is the "minimal" parity symmetric generalized AdS group in eleven
dimensions was argued by Horava \cite{horava} in his recent work on generalized topological Chern-Simons theory in eleven dimensions \cite{private}.}
The two factors of $OSp(1/32,R)_L \times OSp(1/32,R)_R$ correspond to the embedding of left-handed and right-handed spinor representations of $SO(10,2)$ in the fundamental representation of $Sp(32,R)$.  
The contraction to the Poincare superalgebra with central charges proceeds via a diagonal
subsupergroup $OSp(1/32,R)_{L-R}$ which contains the common subgroup $SO(10,1)$
of the two $SO(10,2)$ subgroups. The parity invariant tensor product of the singleton supermultiplets
of the two factors decomposes into an infinite set of "doubleton" supermultiplets of the diagonal
$OSp(1/32,R)_{L-R}$. There is a unique "CPT self-conjugate" doubleton supermultiplet whose
tensor product with itself leads to "massless" supermultiplets. The "CPT self-conjugate" massless
graviton supermultiplet contains fields corresponding to those of 11-dimensional supergravity plus
 additional ones. We conjecture that the doubleton field theory is a superconformal field theory
in ten dimensions that is dual to an AdS phase of M-theory in the same sense as the duality between the $N=4$ super Yang-Mills in $d=4$ and the $IIB$ superstring over $AdS_5 \times S^5$.

\section{Unitary Lowest Weight Representations of Noncompact Groups}             
 \setcounter{equation}{0}

A representation of the lowest weight type of a simple non-compact group  is defined as a representation in which the spectrum of at least one of the generators is bounded from below.                                               
A non-compact simple group G admits unitary lowest weight representations (ULWR)
 if and only if its quotient space G/H       
with respect to its maximal compact subgroup H is an hermitian symmetric space  
\cite{hc}. Thus the complete list of simple               
non-compact groups that have unitary representations of the lowest weight type
follows from the list of irreducible noncompact hermitian symmetric spaces. 
Below we give the complete list of such simple non-compact groups G and their     
maximal compact subgroups H \cite{hc}:\\                                                  
 \newpage
                                                                              
\hspace*{3.0cm}\parbox[t]{5.0cm}{\underline{G}}\parbox[t]{5.0cm}                
{\underline{H}}\\                                                               
\\                                                                              
\hspace*{3.0cm}\parbox[t]{5.0cm}{$SU(p,q)$}\parbox[t]{5.0cm}{$S(U(p)            
\times U(q))$}\\                                                                
\hspace*{3.0cm}\parbox[t]{5.0cm}{$Sp(2n,R)$}\parbox[t]{5.0cm}{$U(n)$}\\         
\hspace*{3.0cm}\parbox[t]{5.0cm}{$SO^*(2n)$}\parbox[t]{5.0cm}{$U(n)$}\\         
\hspace*{3.0cm}\parbox[t]{5.0cm}{$SO(n,2)$}\parbox[t]{5.0cm}{$SO(n)             
\times SO(2)$}\\                                                                 
\hspace*{3.0cm}\parbox[t]{5.0cm}{$E_{6(-14)}$}\parbox[t]{5.0cm}{$SO(10)\times   
U(1)$}\\                                                                        
\hspace*{3.0cm}\parbox[t]{5.0cm}{$E_{7(-25)}$}\parbox[t]{5.0cm}{$E_6            
\times U(1)$}                                                                   
\vskip 0.5 cm                                                                   
                                                                                
The Lie algebra $g$ of a non-compact group G that admits ULWR's has a Jordan    
structure (3-grading) with respect to the Lie algebra $h$ of its maximal        
compact subgroup H.                                                             

\begin{eqnarray}                                                                
g=g^{-1}\oplus g^0\oplus g^{+1}                                                 
\end{eqnarray}                                                                  
where $g^0=h$ and we have the formal commutation relations                      
\begin{eqnarray}                                                                
{[}g^{(m)},g^{(n)}{]}\subseteq g^{(m+n)}\hspace{2.0cm}m,n=\mp 1,0\nonumber          
\end{eqnarray}                                                                  
and $g^{(m)}\equiv 0$ for $\vert m\vert >1.$                                    
\vskip 0.3 cm                                                                   
In \cite{gs} a general method of construction of unitary lowest weight representations (ULWR)
 of non-compact groups  over the Fock space of an arbitrary number of bosonic oscillators was given.    
Some very special cases of this construction had appeared in the physics literature       
previously. For the groups $SU(p,q),\ Sp(2n,R)$ and            
$SO^*(2n)$ the oscillator method yields directly the irreducible ULWR's. For    
the non-compact groups $SO(n,2),\ E_{6(-14)}$ and $E_{7(-25)}$ the naive        
application of the method leads to reducible representations and one then needs to   
project out the irreducible representations.                                    
                                                                                
To construct the ULWR's one realizes the generators of the noncompact group $G$
as bilinears of bosonic oscillators and in the corresponding Fock space ${\cal F}$                                                                  
one chooses  a set of states $|\Omega >$, referred to as the                
"lowest weight vector", which transforms irreducibly under the Lie algebra $h$ of the
maximal compact   subgroup $H$ and which are annihilated by the generators belonging to the       
$g^{-1}$ space. Then by acting on $|\Omega >$ repeatedly with the           
generators belonging to the $g^{+1}$ space one obtains an infinite set of states
\eq
|\Omega >,\ \ g^{+1} |\Omega >,\ \ g^{+1}g^{+1} |\Omega >,...
\en     
which forms the basis of an irreducible unitary lowest weight representation of $g$.
 Irreducibility of the representation of $g$ follows from the       
irreducibility of the representation $ |\Omega >$  under $h$. Generically the    
generators of $g$ are realized as bilinears of bosonic oscillators $a_i(r)$     
satisfying the canonical commutation relations                                  
\begin{equation}
\begin{array}{c}
~\\
{[}a_{i}(r),a^{j}(s){]} = \delta_{i}^{j} \delta_{rs}\hspace{2.0cm}i,j=1,...,n
\\
{[}a_{i}(r),a_{j}(s){]} = 0  \hspace{2.0cm}     r,s=1,...,p
~
\end{array}
\end{equation}

where the upper indices $i,j,k,... $ are the indices in the representation $R$ of $h$ under which the oscillators transform
and $r,s.. =1,2,..p$ label the different sets of oscillators. We denote the creation operators with
upper indices $i,j,..$ which are hermitian conjugates of the annihilation operators.                                                        
\begin{eqnarray}                                                                
a_i(r)^\dagger\equiv a^i(r) \nonumber                
\end{eqnarray}                                                                  
\\                                                                              
Generally, $R$ is the fundamental representation of  
$h$ .
We shall refer to $p$ as the  number  of    colors. The generators are written as color                                     
singlet bilinears but the lowest weight vector $|\Omega >$ and hence the infinite tower
of vectors belonging to the corresponding ULWR can carry color. Depending on the           
non-compact group the minimal number $p$ of colors required to realize the generators 
can be one or two. If  $p_{min}=1$, then the corresponding unitary irreducible representations will be called singletons and if     
$p_{min}=2$, then the corresponding unitary irreducible representations are referred to as doubletons. The   non-compact groups $Sp(2n,R)$ admit singleton unitary irreducible representations \cite{gs,gw,gh,mg89} while the groups   
$SO^*(2n)$ and $SU(n,m)$ admit doubleton unitary irreducible representations \cite{gs,gnw,gm,mgrs,mg89}. The above definition   
of singleton  representations coincides with the definition of singleton  representations of the    
four dimensional $AdS$ group $SO(3,2)$ with the  covering group $Sp(4,R)$ discovered by Dirac         
\cite{pad}. While there exist only two singleton unitary irreducible representations of $Sp(2n,R)$, one finds      
infinitely many doubleton unitary irreducible representations of non-compact groups. The two singleton       
unitary irreducible representations of  $Sp(4,R)$  can be associated with the spin zero and spin   
$\frac{1}{2}$ fields in $AdS$ background. On the other hand the $d=7\ AdS$ group
$SO^*(8)=SO(6,2)$ admits infinitely many doubleton unitary irreducible representations corresponding to      
fields of arbitrarily large spin \cite{gnw}. However, we should note that the doubleton fields are not of     
the form of the most general higher spin fields in $d=7$. Their decomposition   
with respect to the little group $SU(4)\equiv Spin(6)$ in $d=7$ correspond to     
those representations of $SU(4)$ whose Young-Tableaux have only one row \cite{gnw}. Whereas
the general massive higher spin fields correspond to the representations of the 
little group with arbitrary Young-Tableaux.    
                                 
If one replaces the bosonic oscillators in the above construction with fermionic
ones, then one obtains the unitary representations of the compact forms of the  
corresponding groups. Extending the definition given above for singleton and        
doubleton  representations to  compact groups, one finds that $USp(2n)$ admits        
doubleton unitary irreducible representations (finitely many) while the group $SO(2n)$ admits two singleton  unitary irreducible representations \cite{mg90}. The singleton unitary irreducible representations of $SO(2n)$ are the two spinor representations.
    
In general the compact group $USp(2n)$ admits $n$ non-trivial doubleton         
unitary irreducible representations. For $USp(4)$ they are the spinor representation (\underline{4}) and the adjoint representation (\underline{10}).                   
The two singleton (spinor) representations of $SO(2n)$ combine into             
the unique singleton (spinor) representation of $SO(2n+1)$. 
                                                                              
\section{ Unitary Lowest Weight Representations of Noncompact Supergroups}        
 \setcounter{equation}{0}
The extension of the oscillator method to the construction of the ULWR's of     
non-compact supergroups with a Jordan structure with respect to a maximal       
compact subsupergroup was given in \cite{bg}. This method was further developed       
and applied to space-time supergroups and Kaluza-Klein supergravity theories    
in references \cite{gnw,gw,gm,gst}. The general construction of the ULWR's of the noncompact supergroup $OSp(2n/2m,R)$ with the even subgroup $SO(2n)\times Sp(2m,R)$ was studied
in \cite{gh} and the ULWR's of $OSp(2n^*/2m)$ with the even subgroup $SO^*(2n) \times
USp(2m)$ in reference \cite{mgrs}. 
                                                                                
Consider now the  Lie superalgebra $g$ of a non-compact supergroup $G$ that
has a 3-graded structure with respect to a compact subsuperalgebra $g^0$ of maximal rank
\begin{eqnarray}                                                                
g=g^{-1}\oplus g^0\oplus g^{+1}\nonumber                                        
\end{eqnarray}                                                                  
To        
construct the ULWR's of $g$ one realizes it as bilinears of a set of         
superoscillators $\xi_A(\xi^A)$ whose first $m$ components are bosonic and  
the remaining $n$ components are fermionic                                      
\begin{eqnarray}                                                                
\xi_A(r)=\left(\begin{array}{cc}                                                
a_i(r)\\                                                                        
\alpha_\mu(r)\end{array}\right)\ \ \xi^A(r)=\left(\begin{array}{cc}             
a^i(r)\\                                                                        
\alpha^\mu(r)\end{array}\right)\nonumber                                        
\end{eqnarray}                                                                  
\\                                                                              
\hspace*{4.5cm}$i,j=1,...,m\ \ ;\ \ \mu,\nu=1,...,n$\\                          
\\                                                                              
\hspace*{4.5cm}$r,s=1,...,p.$\\                                                 
\\                                                                              
They satisfy the supercommutation relations                                     
\begin{eqnarray}                                                                
{[}\xi_A(r),\ \xi^B(s)\}=\delta_A\,^B\delta_{rs}                                  
\end{eqnarray}                                                                  
where [\ ,\ \} means an anti-commutator for any two fermionic oscillators and a 
commutator otherwise. Furthermore we have                                              

\begin{eqnarray}                                                                
{[}\xi_A(r),\ \xi_B(s)\}=0={[}\xi^A(r),\ \xi^B(s)\}                                 
\end{eqnarray}                                                                  
Typically $g^0$ is the Lie superalgebra $U(m/n)$ and $\xi^A$ and $\xi_A$        
transform in its fundamental representation and its conjugate, respectively.    
Generally the operators belonging to the $g^{-1}$ and $g^{+1}$ spaces are       
realized as di-annihilation and di-creation operators respectively. Consider  now    
 a  lowest weight vector $|\Omega >$, that transforms irreducibly under   
$g^0$ and is annihilated by $g^{-1}$ operators. Acting on $|\Omega>$ with the $g^{+1}$ operators     
repeatedly one generates a ULWR of $g$\\                                        
 
\begin{eqnarray}                                                                
g^{-1}|\Omega >=0\ ,\ \ \ \ g^0|\Omega >=|\Omega'>\hspace{2.0cm}    
\nonumber\\                                                                     
\ \\                                                                            
\{{\rm ULWR}\}\equiv\{|\Omega >,\ g^{+1}|\Omega >,\ g^{+1}g^{+1}        
|\Omega >,...\} 
\end{eqnarray}     
uniquely labelled by $|\Omega>$.                                                             
A supergroup $g$ admits singleton or doubleton unitary irreducible representations       
depending on whether $p_{min}=1$ or $p_{min}=2$, respectively. For example      
the non-compact supergroup $OSp(2n/2m,R)$ with even subgroup $SO(2n)\times      
Sp(2m,R)$ admits singleton representations. The non-compact supergroup          
$OSp(2n^*/2m)$ with even subgroup $SO(2n)^*\times USp(2m)$ admits doubleton     
unitary irreducible representations, as does the supergroup $SU(n,m/p)$ with even subgroup                   
$S(U(n,m)\times U(p))$. There exist only two irreducible singleton              
supermultiplets of the non-compact supergroup $OSp(2n/2m,R)$ \cite{gw,gh}. On the      
other hand, the supergroups $OSp(2n^*/2m)$ and $SU(n,m/p)$ admit infinitely     
many irreducible doubleton supermultiplets \cite{gnw,mgrs,gm}.                                     

Contrary to the situation with noncompact groups,
not all noncompact supergroups that have ULWRs admit a three grading with respect
to a compact subsupergroup of maximal rank. The method of \cite{bg} was generalized to the
case when the noncompact supergroup admits a 5-grading with respect to a 
compact subsupergroup of maximal rank in \cite{mg88}. For example, the superalgebra of  $OSp(2n+1/2m,R)$
admits a 5-grading with respect to its compact subsuperalgebra $U(n/m)$ , but it does not admit 
a three grading with respect to a compact subsuperalgebra of maximal rank for general $n$ and $m$.                                                                                  
All  finite dimensional         
non-compact                                                                     
supergroups do admit a 5-grading (Kantor structure) with respect to a compact   
subsupergroup of maximal rank  \cite{mg88}.                                              
                                                                                
\begin{eqnarray}                                                                
g=g^{-2}\oplus g^{-1}\oplus g^0\oplus g^{+1}\oplus g^{+2} 
\end{eqnarray}                                                                  
                                                                                
The construction of the ULWR's in this more general case proceeds in a similar manner.
One realizes the Lie superalgebra $g$ in terms of superoscillators and chooses a set
of states $|\Omega>$ in the corresponding super Fock space transforming irreducibly
under the maximal compact subsuperalgebra $g^0$ and annihilated by the operators
in $g^{-1}$ subspace.  The by acting on $|\Omega>$
repeatedly by the generators belonging to $g^{+1}$ one generates an infinite set of states that form the basis of an irreducible ULWR of $g$ \cite{mg88}

\subsection{ Massless Supermultiplets of Anti-de Sitter Supergroups}                     
                                                                              
The Poincar\'{e} limit of the singleton representations of the $d=4\  AdS$ group
$SO(3,2)$ is singular \cite{cf1}. However, the tensor product of two singleton unitary irreducible representations of   
$SO(3,2)$ decomposes into an infinite set of massless unitary irreducible representations which do have a    
smooth Poincar\'{e} limit \cite{cf1,gs,mg81}. Similarly, the tensor product of two            
singleton supermultiplets of $N$ extended $AdS$ supergroup $OSp(N/4,R)$         
decomposes                                                                      
into an infinite set of massless supermultiplets which do have a Poincar\'{e}   
limit \cite{mg81,gw,gh,mg89}. The $AdS$ groups $SO(d-1,2)$ in higher dimensions than four that do admit  supersymmetric extensions  have doubleton representations only. The doubleton    
supermultiplets of extended $AdS$ supergroups in $d=5\ (SU(2,2/N))$ and         
$d=7\ (OSp(8^*/2N))$ share the remarkable features of the singleton             
supermultiplets of $d=4$ $AdS$ supergroups. The tensor product of any two       
doubleton                                                                       
supermultiplets decompose into an infinite set of massless supermultiplets      
\cite{gnw,gm,mg89}. In $d=3$ the $AdS$ group $SO(2,2)$ is not simple and is isomorphic to  
$SO(2,1)\times SO(2,1)$. Because of this fact, one has a rich variety of $AdS$  
supergroups in $d=3$ \cite{gst}. Since locally we have $SO(2,1)\approx                 
SL(2,R)\approx SU(1,1)\approx Sp(2,R)$ the $AdS$ supergroups in $d=3$ (and      
hence in $d=2$) admit singleton representations \cite{gst}.                            
                                                                                
Based on the above and other arguments  the following definition of a massless      
representation (or supermultiplet) of an $AdS$ group (or supergroup) 
was proposed in \cite{mg90}:           
                                                                                
{\it A representation (or a supermultiplet) of an $AdS$ group (or supergroup) is
 massless if it occurs in the decomposition of the tensor product of  
two singleton or two doubleton representations (or supermultiplets).}

The tensor product of more than two copies of the singleton or doubleton        
supermultiplets of $AdS$ supergroups decompose into an infinite set of massive  
supermultiplets in the respective dimensions as     
has been amply demonstrated within the Kaluza-Klein supergravity theories       
\cite{gnw,gw,gm}. A noncompact group that admits only
doubleton representations can always be embedded in a larger noncompact group
that admits singleton representations. In such cases the singleton representation of the
larger group decomposes into an infinite tower of doubleton representations    of
the subgroup.                                                    
 
\section{Unitary Supermultiplets of $OSp(1/2m,R)$}
\setcounter{equation}{0}
Simple AdS supergroups with an even subgroup $SO(d-1,2)\times K$ ,where $K$ is a
compact internal symmetry group, exists in $d \leq 7$. Embedding AdS groups 
in simple supergroups in $d >7$ requires enlarging  $SO(d-1,2)$ to a larger simple group
such as $Sp(2^{[d/2]},R)$. For example the supergroup $OSp(1/32,R)$ was proposed as the 
generalized AdS supergroup in d=11 \cite{vhvp}.  As mentioned above
  the noncompact supergroup $OSp(2n+1/2m,R)$  has a five grading with respect to
its compact subsupergroup $U(n/m)$ of maximal rank . For the case $n=0$ the five grading
of $OSp(1/2m,R)$ is with respect to the compact subgroup $U(m)$. In this section we
shall apply the results of \cite{mg88} to construct the unitary lowest weight representations
of $OSp(1/2m,R)$ , in particular, those of $OSp(1/32,R)$. 
Denoting the Lie superalgebra of $OSp(1/2m,R)$ as $g$ we have the following decomposition
in a split basis:
\eqn
g &=& g^{-2} \oplus g^{-1} \oplus g^{0} \oplus g^{+1} \oplus g^{+2} \\ \nn
g&=& L_{ij} \oplus L_i \oplus L_i^j \oplus L^i \oplus L^{ij}
\enn
where $i,j,..=1,2,...m$ . The supersymmetry generators $L^i$ and $L_i$ transform in the fundamental representation and its conjugate under the subalgebra $U(m)$ generated by $L_i^j$, respectively.    
The generators of the grade $\pm 2$ subspaces are symmetric tensors
\eqn
L_{ij} =L_{ji} \\ \nn
L^{ij} = L^{ji}
\enn
We have the following oscillator realization of $OSp(1/2m,R)$ \cite{mg88}:
\eqn
L_i &=& \vec{\psi} \cdot \vec{a}_i + \vec{\psi}^{\dagger} \cdot \vec{b}_i +
\frac{\epsilon}{\sqrt{2}} (\gamma + \gamma^{\dagger} ) c_i  \\ \nn
L_{ij}& =& \vec{a}_i \cdot \vec{b}_j + \vec{a}_j \cdot \vec{b}_i +
\frac{\epsilon}{2} c_i c_j \\ \nn
L^i_j & = & \vec{a}^i \cdot \vec{a}_j + \vec{b}_j \cdot \vec{b}^i +
\frac{\epsilon}{2} (c^ic_j + c_j c^i ) \\ \nn
L^i &=& \vec{\psi}^{\dagger} \cdot \vec{a}^i + \vec{\psi} \cdot \vec{b}^i +
\frac{\epsilon}{\sqrt{2}} (\gamma + \gamma^{\dagger} ) c^i  \\ \nn
L^{ij}& =& \vec{a}^i \cdot \vec{b}^j + \vec{a}^j \cdot \vec{b}^i +
\frac{\epsilon}{2} c^i c^j 
\enn
The bosonic oscillators satisfy the commutation relations

\eqn
{[}c_i, c^j {]} & =& \delta_i^j \\ \nn
{[}a_{i}(r),a^{j}(s){]} &=& \delta_{i}^{j} \delta_{rs} \\ \nn
{[}b_{i}(r),b^{j}(s){]} &=& \delta_{i}^{j} \delta_{rs}\\ \nn
{[}a_{i}(r),a_{j}(s){]} &=& {[}a_{i}(r),b_{j}(s){]} =0 \\ \nn
{[}a_{i}(r),b^{j}(s){]} &=& {[}b_{i}(r),b_{j}(s){]} = 0 
\enn

where $i,j,...=1,2,...m$ ; $r,s,...=1,2,...f$ and $\epsilon =0,1$. 
The non-vanishing  anticommutation relations of the fermionic oscillators $\psi(r)$ and $\gamma$
 are
\eqn
\{ \psi(r), \psi^{\dagger}(s) \} = \delta_{rs} \\ \nn
\{ \gamma, \gamma^{\dagger} \} =1 
\enn

We define the bilinears appearing above as 
$\vec{a}^i \cdot \vec{b}^j = \sum_{r=1}^{f} a^i(r)b^j(r) \, $ etc..
The supercommutation relations of the generators of $OSp(1/2m,R)$ 
take on a very simple form in the above basis:
\eqn
\{ L_i,L_j \}& =& L_{ij} \\ \nn
\{ L^i, L^j \}& =& L^{ij} \\ \nn
\{L_i,L^j \}& = &L_i^j  \\ \nn
{[} L_i,L_j {]} &=& L_{ij} \\ \nn
{[} L^i,L^j {]} &=& L^{ij} \\ \nn
{[} L_{ij}, L^{kl} {]}& =& \delta_j^l L^k_i + \delta_i^k L_j^l + \delta_j^k L_i^l 
+ \delta_i^l L_j^k 
\enn

Choosing $f=0$ and $\epsilon =1$ yields the singleton supermultiplet and $f=1$
and $\epsilon =0$ leads to the massless supermultiplets \cite{mg90}. If the 
number of colors $2f+\epsilon$ is greater than two one gets massive supermultiplets.
The Fock vacuum is defined as the state $|0>$ annihilated by all the bosonic and
fermionic annihilation operators. A lowest weight vector $|\Omega>$ of a unitary  irreducible representation of $OSp(1/2m,R)$
is also a lowest weight vector of its even subgroup $Sp(2m,R)$. Acting on the lowest weight vector $|\Omega>$ by the supersymmetry  generators $L^i$ one generates new lowest weight vectors of
$Sp(2m,R)$. The action of $L^{ij}$ on such a lowest weight vector generates the higher modes
within a unitary irreducible representation of $Sp(2m,R)$. We shall identify the particle states of a unitary  irreducible representation of the "generalized AdS group" $Sp(2m,R)$ with the Fourier modes of a field on that generalized AdS space . As in ref
erences \cite{gnw,gw,gm} these fields will be uniquely
identified  by  the labels of  their lowest weight vectors under the maximal compact subgroup of the AdS group which is $U(m)$ in this case. For labelling fields in AdS space with respect to the $U(m)$ transformation properties of the corresponding \lwv s
  we shall use the Young tableaux notation. 
A field in AdS space whose lowest weight vector transforms in a representation of $U(m)$
with the Young tableau $(n_1,n_2,..,n_m)$ will be labelled as

\eq
\Xi_{(n_1,n_2,...,n_m)_p}
\en
where $p=2f+\epsilon$ is the number of colors and $n_i$ denotes the number of boxes in the $ith $ row
of the Young tableau.
In some suitably chosen units the AdS energy $E$ of this field will be given by
\eq
E=s+ m(f+\frac{\epsilon}{2})= s+\frac{1}{2}mp
\en
  where
 \eq
s= n_1+n_2+ \cdots +n_m
\en
This field is a fermion if $s$ is an odd integer, and a boson when it is an even integer.

\subsection{Unitary Supermultiplets of $OSp(1/32,R)$}

Let us now study the unitary supermultiplets of  $OSp(1/32,R)$ in detail.
 For a single color i.e $f=0$ and $\epsilon=1$ there exists only a single lowest
weight vector of $OSp(1/32,R)$ in the super Fock space , namely the vacuum
state 
\eq
L_i |0>=0 \Longrightarrow L_{ij} |0> =0
\en
Acting on $|0>$ by the generators $L^{ij}$ repeatedly one generates the infinite set
of states corresponding to the Fourier modes of a scalar field $\Phi$ with eight units
of AdS energy. The AdS energy is given by the eigenvalue of the $U(1)$ generator 
$L^i_i$ that determines the 5-grading. Let us  denote an irreducible bosonic 
(fermionic) field in our generalized AdS
space as $\Phi_{d}(E)$ ($\Psi_{d}(E)$) where $d$ labels the $SU(16)$ representation of the
 lowest weight vector of
$Sp(32,R)$ and $E$ stands for the AdS energy. Thus the scalar field defined by the Fock vacuum $|0>$
with a single color is $\Phi_{1}(8)$ :
\eq
|0> \Longrightarrow \Phi_{1}(8)=\Xi_{(0,0,\cdots,0)_1}
\en
By acting with the supersymmetry generator $L^i$ on the lowest weight vector of $OSp(1/32,R)$  we generate
another lowest weight vector of  $Sp(32,R)$ :
\eq
L_{ij} L^{k}|0> = L_{ij} \gamma^{\dagger}c^i |0>=0 
\en
The corresponding AdS field is a fermion $\Psi_{16}(9)$ . 
\eq
\gamma^{\dagger} c^i |0> \Longrightarrow \Psi_{16}(9) =\Xi_{(1,0,\cdots,0)_1}
\en
There are no other lowest weight vectors of $Sp(32,R)$  for a single color. Hence the singleton
supermultiplet consists of a scalar field and a spinor field :
\eq
 \Phi_{1}(8) \oplus \Psi_{16}(9)
\en

By our definition \cite{mg90} the massless supermultiplets are obtained by choosing
two colors i.e $f=1$ and $\epsilon =0$. 
The vacuum vector is always a lowest weight vector of $OSp(1/32,R)$ . Acting on it with supersymmetry
generators we obtain two additional \lwv s of $Sp(32,R)$.  These \lwv s and the corresponding fields
in AdS space are

\eq
|0> \Longrightarrow \Phi_{1}(16) = \Xi_{(0,0, \cdots,0)_2} \nn
\en
\eq
L^i |0> = \psi^{\dagger} a^i |0> \arr \Psi_{16}(17) = \Xi_{(1,0,\cdots,0)_2} \nn
\en
\eq
L^{[i} L^{j]} |0> = a^{[i} b^{j]} |0> \arr \Phi_{120}(18) = \Xi_{1,1,0,\cdots,0)_2}
\en

The \lwv  ~ $a^i|0>$ of $OSp(1/32,R)$ leads to the following massless supermultiplet of fields
\eq
a^i|0> \arr  \Psi_{16}(17)=\Xi_{(1,0,\cdots,0)_2} \nn
 \en 
\eq
L^j a^i |0> = a^j a^i \psi^{\dagger}|0> \arr \Phi_{136} (18)= \Xi_{(2,0,\cdots,0)_2} 
\en
 For two colors the other unitary massless supermultiplets of fields are of the form $(n>1)$.
\eq
\Xi_{(n,0,\cdots,0)_2} \oplus \Xi_{(n+1,0,\cdots,0)_2}
\en
As one increases the number of colors the possible unitary supermultiplets become
much richer and for more than two colors they become massive supermultiplets.
Typically, the shortest   
  supermultiplet for a given number of colors  is obtained by choosing the vacuum $|0>$
as the \lwv  ~ of an extended AdS supergroup \cite{gnw,gw,gm}. We shall refer to these
supermultiplets as the vacuum supermultiplets. The massless vacuum supermultiplets
of AdS supergroups in $d=4$ 
go over to CPT self-conjugate supermultiplets in the Poincare  limit. Therefore,
in analogy with the situation in $d=4$ we may sometimes refer to the vacuum supermultiplets
as CPT self-conjugate supermultiplets in other dimensions.
For three colors the vacuum supermultiplet
consists of the following fields:
\eq
|0> \arr \Phi_{1}(24) = \Xi_{(0,0,\cdots,0)_3}  \nn
\en
\eq
L^i |0> \arr \Psi_{16}(25)=\Xi_{(1,0,\cdots,0)_3} \nn
\en
\eq
L^iL^j|0> \arr \Phi_{120}(26) \oplus \Phi_{136}(26) = \Xi_{(1,1,0,\cdots,0)_3} 
\oplus \Xi_{(2,0,\cdots,0)_3} \nn
\en
\eq
L^iL^jL^k|0> \arr \Psi_{560}(27) \oplus \Psi_{1360}(27) = \Xi_{(1,1,1,0,\cdots)_3}
\oplus \Xi_{(2,1,0,\cdots)_3} 
\en

The vacuum supermultiplet for four colors consist of the following fields:
\eqn
~&\Phi_{1}(32) \oplus \Psi_{16}(33) \oplus
\Phi_{120}(34) \oplus \Phi_{136}(34) \oplus  \\ \nn
~&\Psi_{1360}(35)  \oplus \Psi_{560}(35) 
\oplus \Phi_{5440}(36) \oplus \Phi_{1820}(36) 
\oplus \Phi_{7140}(36)  \\ \nn
~&= \Xi_{(0,\cdots)_4}  \oplus \Xi_{(1,0,\cdots)_4} \oplus \Xi_{(1,1,0,\cdots)_4}
\oplus \Xi_{(2,0,\cdots)_4} \oplus \Xi_{(2,1,0,\cdots)_4} \oplus  \\ \nn
~&\Xi_{(1,1,1,0,\cdots)_4} \oplus \Xi_{(1,1,1,1,0,\cdots)_4} \oplus
\Xi_{(2,1,1,0,\cdots)_4}  \oplus \Xi_{(2,2,0,\cdots)_4} 
\enn
  
By increasing the number of colors  one obtains  more generalized AdS 
supermultiplets. The rules for determining the $U(16)$ labels of the fields
of these supermultiplets are simple and have been given in \cite{gw,gnw,gh,mgrs,mg88}.

\section{$OSp(1/32,R)$ and Physics in Ten and Eleven Dimensions}
\setcounter{equation}{0}
The superalgebra $OSp(1/32,R)$  has a contraction to the 11-dimensional Poincare superalgebra
with two and five form central charges \cite{fre,vhvp} which can be written as \cite{pkt}:
\eq
\{ Q_{A}, Q_{B} \} = (C\Gamma^M)_{AB} P_M + \frac{1}{2} 
(C\Gamma^{MN})_{AB} Z_{MN} + \frac{1}{5!} (C\Gamma^{M_1..M_5})_{AB}
Y_{M_1 ..M_5}  \label{eq:malgebra}
\en
where $\Gamma^M  (M,N=0,1,..,10)$ are the 11 dimensional Dirac matrices, $\Gamma^{M_1..M_p}$  
$ (p=2,5)$ their antisymmetrized products, and $C$ is the charge conjugation matrix.  In the Majorana
representation the 32 component spinors of the Lorentz group $SO(10,1)$ are real and $C$
can be chosen to be $\Gamma^0$. The above superalgebra is to be interpreted 
 simply  as the translation superalgebra with central charges. The Lorentz group $SO(10,1)$ acts
as its automorphism group \cite{pkt}.

Now the group $SO(10,2)$ is also the conformal group in ten dimensions and hence the
group $Sp(32,R)$ can  also be interpreted as a generalized conformal group in ten dimensions.
Generalized spacetimes and superspaces with associated generalized Lorentz and conformal
groups and supergroups were introduced and studied via the theory of Jordan algebras and
Jordan superalgebras \cite{mgjordan}. The conformal Lie algebras $g$ (Lie superalgebras) of Jordan algebras (superalgebras) can be given
a 3-graded decomposition:
\eq
 g= g^{-1} \oplus g^{0} \oplus g^{+1}
\en
where the subspace $g^0 $ is spanned by the generalized Lorentz group generators plus the generator
of dilations  The subspaces $g^{-1}$ and $ g^{+1}$ are spanned by the generators of  translations
and special conformal transformations, respectively. Within this general framework the group $Sp(32,R)$ 
is then the conformal group of the Jordan algebra $J_{16}^R$ of $16 \times 16$ real symmetric
matrices with the generalized Lorentz group $Sl(16,R)$.  It has the decomposition
\eq
Sp(32,R) = K^{\alpha\beta} \oplus ( M^{\alpha}_{\beta} + D ) \oplus T_{\alpha\beta}
\en
where the $T_{\alpha\beta} =T_{\beta\alpha}$ $\alpha, \beta =1,2,..,16$ are translation generators, $K^{\alpha\beta}=K^{\beta\alpha} $  the generators of special conformal transformations , $D$ the dilation generator and $ M^{\alpha}_{\beta}$ are the gene
rators of the generalized Lorentz group
$Sl(16,R)$.
The   $N=1$ supersymmetric extension of this conformal algebra by $Q$ and $S$ type supersymmetry
is simply the superalgebra $OSp(1/32,R)$ with the 5-graded decomposition:
\eq
OSp(1/32,R) = K^{\alpha\beta} \oplus S^{\alpha} \oplus ( M_{\alpha}^{\beta} + D ) \oplus
Q_{\alpha} \oplus  T_{\alpha\beta} 
\en
The conformal supersymmetry generators satisfy the anticommutation relations

\eqn
\{ Q_{\alpha}, Q_{\beta} \} &= &T_{\alpha\beta} \\ \nn
\{S^{\alpha},S^{\beta}\} & = & K^{\alpha\beta} \\ \nn
\{ S^{\alpha},Q_{\beta}\} & = & L^{\alpha}_{\beta} 
\enn

where $L^{\alpha}_{\beta} $ are the generators of $Gl(16,R)= Sl(16,R) + D$.
The decomposition of the generators of $OSp(1/32,R)$ in terms of $SO(9,1)$ covariant operators  
can be read off from the results of  \cite{pkt}:
\eqn
T_{\alpha\beta} &=& (C \Gamma^m \Gamma^+)_{\alpha\beta} P_m 
+ \frac{1}{5!} (C\Gamma^{mnpqr})_{\alpha\beta} Z^+_{mnpqr} \\ \nn  
K^{\alpha\beta}& =& (C\Gamma^m \Gamma^- )^{\alpha\beta} K_m +
\frac{1}{5!} (C\Gamma^{mnpqr})^{\alpha\beta} Z^-_{mnpqr} \\ \nn
L^{\alpha}_{\beta} &= &\delta^{\alpha}_{\beta} D + \frac{1}{2} (\Gamma^{mn} )^{\alpha}_{\beta}
M_{mn} + \frac{1}{4!} (\Gamma^{mnpq})^{\alpha}_{\beta} Y_{mnpq}  
\enn
where $P_m, K_m, D, M_{mn}$ are the usual generators of ten dimensional translations, special conformal transformations, dilations and Lorentz transformations. The $Z^{\pm} $ are the selfdual and antiselfdual 5-form charge generators and Y is a 4-form cha
rge generator. Note that these generators are not central charges of the ten dimensional conformal superalgebra. When restricted to maximal
parabolic subalgebras  $g^0 \oplus g^{\pm 1/2} \oplus g^{\pm 1}$ the generators $Z^{\pm}$ can be interpreted as central charges.

In the last reference of \cite{pkt} the algebras  \ref{eq:malgebra} and $5-6$ were referred to as the M-theory superalgebra and  the membrane superalgebra , respectively, and the differences between them were stressed. The fact that both the uncontracted 
form of  \ref{eq:malgebra} and the membrane superalgebra turn out to be $OSp(1/32,R)$ was 
considered as a coincidence. As we shall explain below the fact that these two algebras are related 
via $ OSp(1/32,R)$ is not a coincidence from our point of view and is the well-established connection
between AdS supergroups in $d+1$ dimensions and the conformal supergroups in $ d $ dimensions. 

Now the M-theory superalgebra \ref{eq:malgebra} is obtained from the supergroup $OSp(1/32,R)$ via an In\"{o}n\"{u} -Wigner type contraction. 
An intermediate step in this
contraction is to decompose the algebra $OSp(1/32,R)$ with respect to the $SO(10,2)$ subgroup
of $Sp(32,R)$. Interpreting this $SO(10,2)$ as the AdS group in 11 dimensions one then takes
the standard In\"{o}n\"{u}-Wigner contraction to the Poincare group.
This embedding of AdS group $SO(10,2)$ in $Sp(32,R)$ is achieved by identifying the fundamental representation of $Sp(32,R)$ with the spinor representation of $SO(10,2)$. The fundamental representation
of $Sp(32,R)$ decomposes as $16\oplus \overline{16}$ under its maximal compact subgroup $SU(16)\times U(1)$.
The fundamental representation $16$ of $SU(16)$ is then identified with the Weyl spinor of 
the $SO(10)$ subgroup of $SO(10,2)$. In contracting to the Poincare group, the fundamental
representation of $Sp(32,R)$ goes over to the spinor representation of the Lorentz group $SO(10,1)$
in eleven dimensions. 

If $OSp(1/32,R)$ is a generalized AdS symmetry of M-theory then among its massless supermultiplets 
there must be one whose contraction  includes the fields of 11 dimensional supergravity.
A study of the massless supermultiplets given in the previous section makes evident that
there do not exist such an irreducible supermultiplet. Thus $OSp(1/32,R)$ can not naively be 
identified with the generalized
AdS supersymmetry of M-theory.  This is actually not surprising since the embedding of
$SO(10,2)$ in $Sp(32,R)$ is not parity invariant. As was shown in \cite{hw,horava} M-theory and
its low energy effective theory in eleven dimensions are parity invariant. In embedding
$SO(10,2)$ in $Sp(32,R)$ we identified its spinor representation with the fundamental representation
of $Sp(32,R)$. However, this embedding is not unique as $SO(10,2)$ has two spinor representations
of dimension 32, i.e the left handed and right handed spinors that are related by parity. We shall
denote the superalgebra in which the the fundamental representation of $Sp(32,R)$ is identified
with the left-handed (right-handed) spinor representation of $SO(10,2)$ as $OSp(1/32,R)_L$
($OSp(1/32,R)_R$). 
If we are to have  parity invariant supermultiplets we must embed the full 64 dimensional
Dirac spinor of $SO(10,2)$ in a larger supergroup. The minimal supergroup satisfying this requirement
is $OSp(1/32,R)_L \times OSp(1/32,R)_R$ with the proviso that one restricts oneself to parity invariant representations \cite{horava}.  In the contraction of $OSp(1/32,R)_L \times OSp(1/32,R)_R$ to obtain the M-theory algebra one identifies the two spinor
 representations hence breaking the $SO(10,2)_L
\times SO(10,2)_R$ symmetry down to the diagonal $SO(10,1)$ subgroup. The situation is rather similar to the situation in 2+1 dimensions where the AdS group $SO(2,2)$ decomposes as
\eq
SO(2,2) =  SO(2,1)_+ \times SO(2,1)_- = Sp(2,R)_+\times Sp(2,R)_-
\en
The  spinorial generators transform in the  representations $(\frac{1}{2},0)$ and  $(0,\frac{1}{2})$
of the AdS group.
The complete list of AdS supergroups in $d=3$ was given in \cite{gst}.
For our discussion and  comparison with eleven dimensions we shall consider  AdS supergroups  of the form 
\eq
OSp(p/2,R)_+ \times OSp(q/2,R)_-
\en
There exist a one parameter family of Chern-Simons supergravity actions of the form \cite{at,ew88}
\eq
S= a_+ S_+ + a_- S_-
\en
where $S_+$ and $S_-$ are the C-S actions for $OSp(p/2,R)_+$ and $ OSp(q/2,R)_-$ 
,respectively, and $a_{\pm}$ are some real constants. Not all these AdS supergravity
theories in $d=3$ have a Poincare limit. Those AdS gravity and supergravity
theories that do not have a Poincare limit are generally referred to as "exotic"
\cite{ew88,at}. To obtain an AdS supergravity which admits a Poincare limit
one needs to take different forms of the two OSp factors such that they differ by an
overall sign in the anticommutators of the supersymmetry generators. 
One crucial difference between the three dimensional AdS supergroups 
$OSp(p/2,R)_+\times OSp(q/2,R)_-$ and the eleven dimensional generalized
AdS supergroups of the form $OSp(1/32,R)_L \times OSp(1/32,R)_R$ is that
the spinorial representations of the supersymmetry charges in the two factors are
isomorphic in three dimensions and non-isomorphic in eleven dimensions.

Consider now the chiral component $OSp(1/32,R)_L$ of the eleven dimensional 
generalized AdS supergroup. It admits a singleton supermultiplet which consists
of a scalar field $\Phi$ and a left-handed spinor $\Psi_L$ of $SO(10,2)$. 
Since it is a singleton supermultiplet we expect its quantum field theory to be 
a conformally invariant theory in ten dimensions.  The massless supermultiplets
are obtained by tensoring two singleton supermultiplets and decomposing them
into irreducible supermultiplets. 
By decomposing the maximal compact subgroup $U(16)$ representations
of the lowest weight vectors with respect to the $SO(10)\times U(1)$ subgroup
of $SO(10,2)$ one can determine the field content of the supermultiplets in
$ AdS_{11}$.  The massless vacuum
supermultiplet of $OSp(1/32,R)$ include a scalar field $\Phi$, a left-handed spinor $\Psi_L$
and an anti-symmetric tensor field $A_{\mu\nu\rho}(\mu,\nu,..=0,1,...10)$ and does not include the graviton. The massless graviton supermultiplet of $OSp(1/32,R)_L$ include the graviton $g_{\mu\nu}$, a gravitino $\Psi_{\mu}$  a spinorial field whose lowes
t weight vector transforms in the $\underline{672}$ representation of the little group $SO(10)$,
and two bosonic fields whose lowest weight vectors transform in the $\underline{1050}$ and
$\underline{2772}$ of $SO(10)$.  To obtain an irreducible vacuum supermultiplet that contains
the graviton  one needs to consider four colors, which by our definition would correspond to 
a massive supermultiplet. The other chiral component $OSp(1/32,R)_R$ has similar
supermultiplets with the left handed spinorial indices replaced by right handed spinorial indices.

Now if we consider the full generalized AdS supergroup $OSp(1/32,R)_L\times OSp(1/32,R)_R$
then to construct  supersymmetric AdS quantum field theories that are parity invariant we need to consider left-right symmetric supermultiplets. In particular, we would like to construct an AdS
supergravity that in the Poincare limit leads to the 11-dimensional supergravity theory of \cite{cjs}.
In the contraction to the 11 dimensional Poincare superalgebra we will consider the diagonal
subgroup $OSp(1/32,R)_{L-R}$ of $OSp(1/32,R)_L \times OSp(1/32,R)_R$. Here we should
stress the important point that the groups $SO(10,2)_L$ and $SO(10,2)_R$ intersect within
their $SO(10,1)$ subgroups inside $OSp(1/32,R)_{L-R}$.  The simplest parity invariant
irreducible  representation of \slr  ~is  the tensor product of the  singleton supermultiplets 
 of $OSp(1/32,R)_L$ and $OSp(1/32,R)_R$ . This tensor product however is infinitely reducible with respect to the diagonal subsupergroup $OSp(1/32,R)_{L-R}$. The irreducible supermultiplets contained in this tensor product are the  supermultiplets of $OSp
(1/32,R)_{L-R}$ for two colors
constructed in the previous section. ( For $OSp(1/32,R)_L$ and $OSp(1/32,R)_R$ they would be the massless multiplets.) However, for the diagonal supergroup $OSp(1/32,R)_{L-R}$ they are the shortest possible supermultiplets consisting of  parity invariant 
irreducible ones
plus those that come in parity conjugate pairs.  Therefore by an abuse of terminology we shall refer to these supermultiplets as "doubleton supermultiplets " of $OSp(1/32,R)_{L-R}$ 
and the supermultiplets one obtains by tensoring two doubletons as "massless " supermultiplets.
 It is easy to show that the 
decomposition of the  parity invariant irreducible supermultiplets of 
$OSp(1/32,R)_L \times OSp(1/32,R)_R$ with respect to $OSp(1/32,R)_{L-R}$ yields supermultiplets
corresponding to an even number of colors. In other words they consist of    doubletons  and those
supermultiplets  that can be obtained by tensoring doubleton supermultiplets 

Based on the structure of the supermultiplets for two colors and the knowledge of currently 
known supersymmetric theories in ten and eleven dimensions \cite{ss} I expect  the doubleton
supermultiplets  not to have a Poincare limit in eleven dimensions and hence their field
theories must be conformally invariant theories on the boundary of the $AdS_{11}$ which
we can identify with the ten dimensional Minkowski space . Indeed,
there exist conformally invariant supergravity theories with and without matter couplings in ten dimensions \cite{brw,ss}.  It will be important to study their connection to the doubleton supermultiplets
given above. Of all the supermultiplets for two colors,  the vacuum doubleton supermultiplet consisting of
a scalar field $\Phi$ , a spinor field $\Psi$ and an antisymmetric tensor field $A_{\mu\nu\rho}$
($\mu, \nu,..=0,1,..9$) seems to be of central importance as will be explained shortly. 
 
Denoting a bosonic and a fermionic field corresponding to a lowest weight vector transforming
in the representation $r$ of the little group $SO(10)$ as $B_{(r)}$ and $F_{(r)}$ , respectively,
the massless vacuum supermultiplet (p=4) of $OSp(1/32,R)_{L-R}$ has the decomposition:
\eqn
\Phi_{1}(32) &=& B_{1} \\ \nn
\Psi_{16}(33)&=& F_{16} \\ \nn
\Phi_{120}(34) &=& B_{120} \\ \nn
\Phi_{136}(34) &=& B_{10} + B_{126} \\ \nn
\Psi_{560}(35) &=& F_{560} \\ \nn
\Psi_{1360}(35) &=& F_{\overline{16}} + F_{\overline{144}} + F_{\overline{1200}}\\ \nn
\Phi_{1820}(36)&=& B_{770} + B_{1050} \\ \nn
\Phi_{5440}(36)&=& B_{1}+B_{54}+B_{210}+B_{1050}+B_{4125} \\ \nn
\Phi_{7140}(36)&=& B_{45}+B_{210}+B_{945}+B_{5940} 
\enn
Since it contains the fields that in the Poincare limit go over to the fields of eleven
dimensional supergravity we shall refer to it as the massless $AdS_{11}$ graviton supermultiplet.
In addition to the fields of 11 dimensional Poincare supergravity, the above
supermultiplet contains  extra bosonic and fermionic fields. Furthermore, as is typical
in AdS supersymmetric theories there is a mismatch of bosonic and fermionic degrees of
freedom. One ought to keep in mind however that in taking the Poincare limit many
 of the generators of $OSp(1/32/R)_{L-R}$
will become central charges and some  degrees of freedom of the fields 
will become gauge degrees of freedom.
In any case, what is evident  is that an AdS supergravity theory in
eleven dimensions would require additional fields than those of ordinary Poincare supergravity.
This may explain why the attempts to construct a gauged version of 11 dimensional supergravity
with 128 bosonic and 128 fermionic degrees of freedom have so far produced negative
results \cite{bdhs}.
We should also note that as in $d=2+1$ dimensions it may be possible to construct exotic 
AdS supergravity theories in $d=10+1$ dimensions that have no Poincare limit. 

Assuming that the $AdS_{11}$ supergravity corresponding to the above supermultiplet exists, then
it is related to the  superconformal  field theory of the vacuum doubleton supermultiplet in ten dimensions
  as the gauged $N=8$ supergravity in $d=5$ theory is related to the $N=4$ super Yang-Mills
theory in $d=4$ which is  conformally invariant \cite{gm,mgnm2}. In \cite{gm}
the entire spectrum of the $S^5$ compactification of $IIB$ supergravity in $d=10$ 
were fitted into massless and massive vacuum supermultiplets of the $N=8$ AdS superalgebra
$SU(2,2/4)$. The doubleton supermultiplet of $SU(2,2/4)$ is nothing but the $N=4$ Yang-Mills
supermultiplet living on the boundary of $AdS_5$ which is identified with the $d=4$ Minkowski
space. The entire spectrum of $IIB$ can be obtained by tensoring the doubleton supermultiplet
with itself repeatedly and restricting to the CPT self-conjugate color singlet supermultiplets.
As explained in \cite{mgdm} this is in complete parallel with the recent conjecture of Maldacena that the $N=4$ supersymmetric Yang-Mills theory with the gauge group $SU(n)$ is equivalent to the $IIB$
superstring on $AdS_5 \times S^5$ in the large $n$ limit. 

If M-theory admits a phase whose low energy effective theory is an  $AdS_{11}$ supergravity
theory corresponding to the above supermultiplet then I expect M-theory in that phase to be
dual to the "CPT self-conjugate" doubleton field theory in ten dimensions. In particular the entire spectrum of the AdS phase of M-theory , massless as well as massive, should then be obtainable by tensoring  the doubleton supermultiplet with itself repea
tedly. In other words, the doubleton
field theory would then be the holographic relativistic quantum field theory underlying
M-theory in the sense of \cite{gth,ls,ew98}.

As in $d=2+1$ dimensions there may exist AdS supergravity theories based on chiral
superalgebras of the form $OSp(1/32,R)_L\times OSp(1/32,R)_L$ which do not 
have any Poincare limit in eleven dimensions. However, they may have a Poincare
limit in ten dimensions which is related to the type IIB supergravity and/or conformal
supergravity. \footnote{ We should note that the topological Chern-Simons supergravity theories
in eleven dimensions based on the non-unitary adjoint representations of  $OSp(1/32,R)$ and $OSp(1/32,R)\times OSp(1/32,R)$ were studied in  \cite{ac,tz,horava}. In the first paper of
\cite{tz} it was shown that there is a contraction of $OSp(1/32,R)$ that leads to a non-standard
supergravity theory with a Poincare and parity invariant gravity sector. The connection of
 such non-standard supergravity theories  to Cremmer-Julia-Scherk supergravity , if any, is not known \cite{tznote}.}  
 Both the left-right
symmetric superalgebra $OSp(1/32,R)_L\times OSp(1/32,R)_R$   and the chiral superalgebra
$OSp(1/32,R)_L\times OSp(1/32,R)_L$ can be embedded in the simple superalgebra
$OSp(1/64,R)$ which had been discussed in connection with the 11-dimensional supergravity
theory \cite{fre,vhvp}and more recently for unification of various duality groups in string/M-theory
\cite{bars}. The unitary supermultiplets of $OSp(1/64,R)$ can be written down easily using the
results of section 4 above and those of \cite{mg89}. The physical interpretation of these
supermultiplets and their relation to M-theory as well as the construction of ten dimensional
singleton and doubleton superconformal field theories will be discussed elsewhere
\cite{mg98}. 

{\bf Acknowledgements:} I would like to thank D. Minic for discussions.


\begin{thebibliography}{99}
\bibitem{private} I would like to thank P. Horava for a discussion on this point.
\end{thebibliography}
\end{document}